\documentclass[sigconf]{acmart}
\usepackage{graphicx} % Required for inserting images
\usepackage{url}
\usepackage{hyperref}
\usepackage{color}
\usepackage{microtype}
\usepackage{subfigure}
\usepackage{booktabs} % for professional tables
\usepackage{subcaption}
\usepackage{multirow}
\usepackage{amsmath}
\usepackage{subcaption}
\usepackage{mathtools}
\usepackage{amsthm}
\usepackage{float}
\usepackage{algorithm}
\usepackage{algorithmic}
\usepackage{enumitem}
\usepackage{relsize}
\usepackage{multirow}
\newtheorem{thm}{Theorem}
\newtheorem{lem}{Lemma}
\newtheorem{prop}[thm]{Proposition}
\newtheorem{srule}[lem]{Scaling Rule}

\title{Two-dimensional Sparse Parallelism for Large Scale Deep Learning Recommendation Model Training}
\author[Zhang et al.]{%
Xin Zhang$^*$ \quad Quanyu Zhu$^*$ \quad Liangbei Xu$^*$ \quad Zain Huda$^*$  \quad
{Wang Zhou} \quad {Jin Fang} \\
{Dennis van der Staay}   \quad {Yuxi Hu} \quad
{Jade Nie} \quad {Jiyan Yang}$^\dagger$ \quad {Chunzhi Yang}$^\dagger$ \\
Meta, Inc.\\
\texttt{\{qyz, liangbei, xinzhang5, zainhuda, wangzhou, fangjin\}@meta.com}\\
\texttt{\{dstaay, yuxihu, qnie, chocjy, zorror\}@meta.com}\\
$^*$Equal Contributors; $^\dagger$Corresponding Authors
}

\date{January 2025}

\begin{document}
\begin{abstract}
The increasing complexity of deep learning recommendation models (DLRM) has led to a growing need for large-scale distributed systems that can efficiently train vast amounts of data. 
In DLRM, the sparse embedding table is a crucial component for managing sparse categorical features. Typically, these tables in industrial DLRMs contain trillions of parameters, necessitating model parallelism strategies to address memory constraints.
However, as training systems expand with massive GPUs, the traditional fully parallelism strategies for embedding table post significant scalability challenges, including imbalance and straggler issues, intensive lookup communication, and heavy embedding activation memory. 
To overcome these limitations, we propose a novel two-dimensional sparse parallelism approach. Rather than fully sharding tables across all GPUs, our solution introduces data parallelism on top of model parallelism. This enables efficient all-to-all communication and reduces peak memory consumption. Additionally, we have developed the momentum-scaled row-wise AdaGrad algorithm to mitigate performance losses associated with the shift in training paradigms.
Our extensive experiments demonstrate that the proposed approach significantly enhances training efficiency while maintaining model performance parity. It achieves nearly linear training speed scaling up to 4K GPUs, setting a new state-of-the-art benchmark for recommendation model training.

\end{abstract}
\maketitle

\section{Introduction}\label{Sec.intro}
% !TEX root = ../main.tex

The advent of big data and recent advancements in artificial intelligence have led to the widespread adoption of deep learning recommendation models (DLRM) across various industrial fields and applications, including advertising, search and social media, etc. \cite{naumov2019deep,zhang2019deep,elkahky2015multi,covington2016deep,lyu2020deep}.
By providing users with personalized content that aligns with their preferences and interests, DLRM has proven instrumental in driving revenue growth. Recently, studies \cite{zeng2024interformer,liang2025external,zhaiactions,ardalani2022understanding,zhang2024scaling,zhang2024wukong,zhang2022dhen} have consistently demonstrated that scaling up model complexities can yield substantial performance gains, resulting in enhanced user engagement, increased conversions, and ultimately, improved return on investment.
To support the development of large-scale recommendation models, it is essential to scale up the training system with massive GPU capabilities, enabling faster and more efficient processing of vast amounts of data \cite{mudigere2022software,zhao2023pytorch,xu2020automatic,cai2024distributed,naumov2020deep}.

In the realm of recommendation systems, sparse categorical feature exists as a unique and crucial feature type \cite{kang2021learning,cheng2016wide,jha2024mem,singh2024better}. These features, such as user IDs or item IDs, play a vital role in capturing essential information about users' preferences and behavior. To effectively utilize these features, DLRM employs the embedding learning technique, which stores comprehensive embedding tables and map sparse categorical IDs to dense embeddings through the row-based look-up \cite{naumov2019deep,tsang2023clustering,coleman2023unified}. 
Due to the heavy memory consumption, the embedding tables are sharded across GPUs using model parallelism within a distributed training paradigm \cite{zhao2020distributed,liu2024embedding}.
However, as models and training systems continue scaling up, the traditional embedding tables' parallelism methods introduce substantial scalability challenges:
\begin{enumerate}[leftmargin=14pt, itemindent=5pt]
     \item \textit{Imbalance and straggler issue}: It becomes hard for naive model parallelism strategies to achieve balanced sharding across massive number of GPUs. Some GPUs can have more workload on embedding computation and communications than the others, which slow down global training.
    \item \textit{Intensive look-up communication}: With the distributed embedding table sharding, the GPUs need the global all-to-all embedding look-up to obtain the completed embedding representations \cite{mudigere2022software}. The all-to-all communication becomes costly as the number of GPUs grows. 
    \item \textit{Heavy embedding activation memory}: The GPUs store the local look-up embedding activations of the global data batch. Given a fixed batch size per GPU, the memory cost of the activation would grow linearly at the scale of the number of GPUs.
\end{enumerate}
These memory and communication overheads would significantly impact model performance and training efficiency.

To tackle the challenges posed above by large-scale distributed training systems, we propose a novel parallelism strategy called \textit{two-dimensional sparse parallelism} (2D sparse parallelism) for embedding tables.  Our key contributions are as follows: 
\begin{itemize}[leftmargin=10pt, itemindent=5pt]
    \item We first analyze the system limitations of existing model parallelism approaches for embedding tables under a large-scale distributed training system, thereby gaining a comprehensive understanding of the performance bottlenecks. We then present the system design and training algorithm of our 2D sparse parallelism strategy. Lastly, we elaborate on how our design overcomes the identified limitations and highlight its advantages, providing a thorough evaluation of its effectiveness.
    \item The row-wise AdaGrad algorithm is a widely adopted training algorithm for embedding tables in DLRM. However, we observe that incorporating 2D sparse parallelism with the standard row-wise AdaGrad leads to an unexpected loss in model performance. Through quantitative analysis, we identify the root cause of this issue as an effective learning rate reduction. To mitigate this issue, we develop a novel moment-scaled row-wise AdaGrad algorithm and provide theoretical justification for its effectiveness.
    \item Finally, we validate the efficacy of our proposed 2D sparse parallelism and moment-scaled row-wise AdaGrad algorithm through an extensive series of experiments. The results demonstrate that our method significantly enhances training efficiency by facilitating fast all-to-all communication and reducing peak memory consumption, while maintaining parity in model performance. By adopting the proposed 2D sparse parallelism, we scale the training system up to $4K$ GPUs and achieve almost linear training speed scaling up. The findings confirm the effectiveness of the proposed approach in addressing the challenges of large-scale distributed training systems and refresh the state-of-the-art record for recommendation model training.
\end{itemize}
The rest of the paper is organized as follows. 
In Section~\ref{Sec.prelim}, we provide a comprehensive overview of the preliminaries, including DLRM and embedding tables' parallelism strategies.
We then present our proposed 2D sparse parallelism in Section~\ref{Sec.2D method}, which consists of system design (Section~\ref{Sec.2D system}) and algorithm development (Section~\ref{Sec.2D algorithm}).
Next, we evaluate the efficiency of our method with experimental studies and report the results in Section~\ref{Sec.experiments}.
We conclude the paper in Section~\ref{Sec.conclusion}.
% In the end, we discuss the deployment considerations and strategies in Section~\ref{Sec.2D optim}, and conclude the paper in Section~\ref{Sec.conclusion}

\section{Preliminaries}\label{Sec.prelim}
% !TEX root = ../main.tex

In this section, we begin with a background review of distributed DLRM systems and the use of model parallelism for embedding tables. We then discuss the limitations of traditional model parallelism approaches for handling embedding tables, which motivated the development of our 2D sparse parallelism.

\subsection{Distributed DLRM and Embedding Table Parallelism}

% -> in indurial enviroment DLRM we have lots of data and sparse tables -> have to do the model parallelsim -> how model parallelsim works for sparse
% \begin{figure}
%     \centering
%     \includegraphics[width=\linewidth]{Figures/fig_system.pdf}
%     \vspace{-.2in}
%     \caption{The common distributed system for large-scale DLRM. The embedding tables are sharded in GPUs with model parallelism while the deep neural network parameters are data parallelism across GPUs.}
%     \label{fig:DLRM illustration}
% \end{figure}

DLRMs are designed to predict user interests and behaviors by leveraging the user and product information.
Typically, users and products are characterized by a combination of continuous float features and sparse categorical features, which serve as the input features for DLRMs.
\begin{figure}
  \begin{center}
    \includegraphics[width=0.8\linewidth]{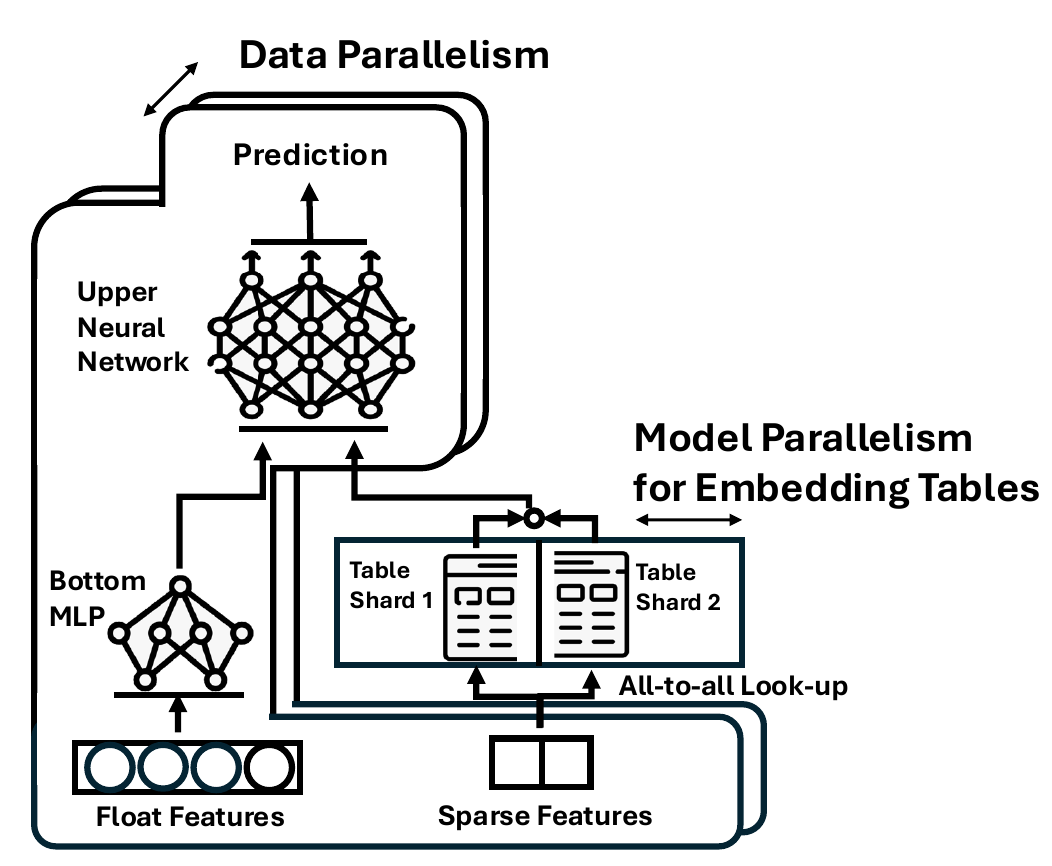}
  \end{center}
  \caption{The common distributed system for large-scale DLRM. The embedding tables are sharded in GPUs with model parallelism while the deep neural network parameters are data parallelism across GPUs.}
    \label{fig:DLRM illustration}
    \vspace{-.1in}
\end{figure}
Similar to traditional models in computer vision and natural language processing, DLRMs process continuous float features using the multi-layer perceptrons (MLPs) to transform them into high-dimensional embedding vectors. 
For sparse categorical features, modern DLRMs employ embedding learning techniques to generate their embedding representations. Specifically, each sparse categorical feature is associated with an individual embedding table, where each row corresponds to different IDs within the categorical feature, and each column corresponds to the embedding dimensions. During the training, DLRMs will map the categorical IDs to the embedding vectors by looking up entries in embedding tables.
These embedding vectors are then combined with the processed dense embedding vectors. 
The aggregated embeddings are passed to the upper-level neural network for the prediction. We provide the illustration of a DLRM in Figure~\ref{fig:DLRM illustration}.

Industrial DLRMs usually include hundreds or even thousands of sparse categorical features, many of which have millions of unique IDs. This results in substantial memory requirements.
For instance, \cite{liang2025external} proposed the external large foundation model framework for online ads recommendation, in which the largest DLRM contains maximally 3.2 trillion parameters, with over 98\% of these parameters residing in the embedding tables. These tables require more than 5TB of memory, far exceeding the capacity of a single GPU.
To train model of such size, modern DLRMs employ model parallelism to \textit{fully} distribute embedding tables across multiple GPUs, as discussed in \cite{mudigere2022software}. Common embedding table parallelism strategies include table-wise, row-wise, column-wise, and combinations of them.
In the forward pass, each GPU performs embedding look-ups for the IDs, including those from mini-batches assigned to other GPUs, from the sharded tables it hosts.
The retrieved embeddings are then transferred to the appropriate GPUs through all-to-all communication, ensuring that each GPU receives the embeddings for the IDs present in its own mini-batch.
In the backward pass, gradients corresponding to each table row are sent to the relevant GPUs via all-to-all communication. It's important to note that only the table rows with IDs appearing in the batch will have associated gradients.
To optimize memory usage, the embedding tables employ a fused backward and optimization process. The optimization step is executed immediately after the gradients are obtained during the backward pass, allowing for the elimination of gradients and freeing up memory \cite{khudia2021fbgemm}.

\subsection{Challenges under Massive GPU Training}

\begin{figure}
    \centering
    \includegraphics[width=\linewidth]{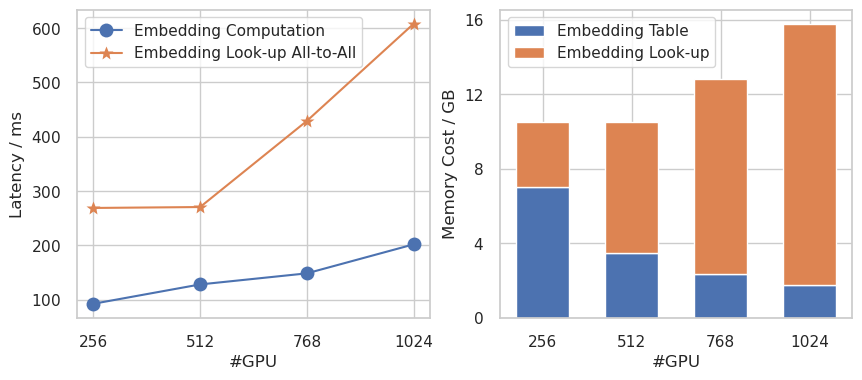}
    \caption{The latency and memory bottlenecks caused by embedding tables with full model parallelism. Left is the maximum sparse all-to-all communication and computation latency; and right is the maximum GPU memory consumption related with embeddings.}
    \label{fig:scaling_bottleneck}
    \vspace{-.2in}
\end{figure}

Recent studies have verified that scaling up model architectures, such as stacking hierarchical neural network layers \cite{zhang2022dhen,zhaiactions} or incorporating complex sequential models \cite{zeng2024interformer}, can significantly enhance model performance.   
As model complexity increases rapidly, distributed parallel computing becomes essential to maintain a reasonable training speed.
While models are scaled up, the sparse categorical features and the size of embedding tables typically remain unchanged.
Although traditional model parallelism can still reduce the memory cost of embedding tables on each GPU, it faces training inefficiencies due to the sharding imbalance and straggler issues, which further causes the intensive look-up communication and significant memory usage for embedding look-up activations.
Here we take an external large foundation model from \cite{liang2025external} as an illustration. 
The model has more than 4000 embedding tables with about 2TB memory size. 
We show the communication latency and memory bottleneck of this model in the left-hand side of Figure~\ref{fig:scaling_bottleneck}.
As the number of GPUs is scaled up, it can be observed that the embedding computation latency increases gradually, from 100ms to 200ms. However, the communication all-to-all latency increases at a much faster rate than the computation latency. Specifically, when training with $1K$ GPUs, the all-to-all latency exceeds 600ms. This implies that the all-to-all communication becomes a bottleneck, hindering the potential of GPU scaling.
% In the left of Figure~\ref{fig:scaling_bottleneck}, we show the latency with respect to the GPU number scaling.
% It can be observed that increasing the number of GPUs from 256 to 1024 leads to a significant rise in both the maximum embedding computation time and the maximum all-to-all communication latency across ranks. This increase is attributed to sharding imbalance, which consequently slows down the training speed. 
On the other hand, as shown in the right-hand side of Figure~\ref{fig:scaling_bottleneck}, the memory cost of embedding look-ups increases significantly: when training with 256 GPUs, the embedding look-ups consumes less than 4GB; however, it grows to almost 15GB when adopting $1K$ GPUs for training. 
Although the per-GPU embedding table size is reduced, the memory benefits are limited as the number of GPUs increase. The intensive memory costs associated with embedding look-up lead to out-of-memory issues, making additional GPU scaling infeasible.
% We observe that embedding all-to-all communication latency increase almost exponentially when we scale up the number of trainers from 256 to 1024 (see the left figure in Figure~\ref{fig:scaling_bottleneck}). This significantly slows down the training speed. 
% Also the out-of-memory failure are frequently hit due to extensive activation memory cost as illustrated in the right figure in Figure~\ref{fig:scaling_bottleneck}. 
The literature \cite{liu2024embedding,mudigere2022software,zha2022autoshard,zha2023pre,zha2022dreamshard} contains extensive research efforts on improving training efficiency in terms of all-to-all communication latency and memory cost. However, these studies are limited to small GPU scale and cannot address the challenges brought by massive GPU training (i.e. $O(1K)$ GPUs and beyond).

% \begin{figure}
%     \centering
%     \includegraphics[width=0.95\linewidth]{Figures/fm_kernel_breakdown.png}
%     \caption{All-to-all communication and computation latency vs \#GPU for exLarge foundation model with traditional full model parallelism approach.}
%     \label{fig:fm_breakdown_vs_num_gpu}
% \end{figure}

% \begin{figure}
%     \centering
%     \includegraphics[width=0.95\linewidth]{Figures/memory_scaling.pdf}
%     \vspace{-.1in}
%     \caption{Maximum GPU memory cost vs \#GPU for exLarge foundation model with traditional full model parallelism approach.}
%     % \caption{Per-GPU memory cost vs \#GPU for exLarge foundation model with traditional fully sharding approach.}
%     \label{fig:memory_scaling}
% \end{figure}

\section{Two-dimensional sparse parallelism} \label{Sec.2D method}

In this section, we introduce our proposed solution to aforementioned technical challenges under large-scale DLRM training: two-dimensional (2D) sparse parallelism. Section~\ref{Sec.2D system} outlines the system architecture of 2D sparse parallelism. Subsequently, Section~\ref{Sec.2D algorithm} presents a tailored training algorithm for 2D sparse parallelism, termed momentum-scale row-wise AdaGrad. 
% Furthermore, we discuss the potential overheads of our 2D sparse parallelism approach and offer corresponding mitigation strategies in Section~\ref{Sec.2D optim}.

\subsection{System Design} \label{Sec.2D system}
% !TEX root = ../main.tex

\begin{figure*}
    \centering
    \includegraphics[width=0.98\linewidth]{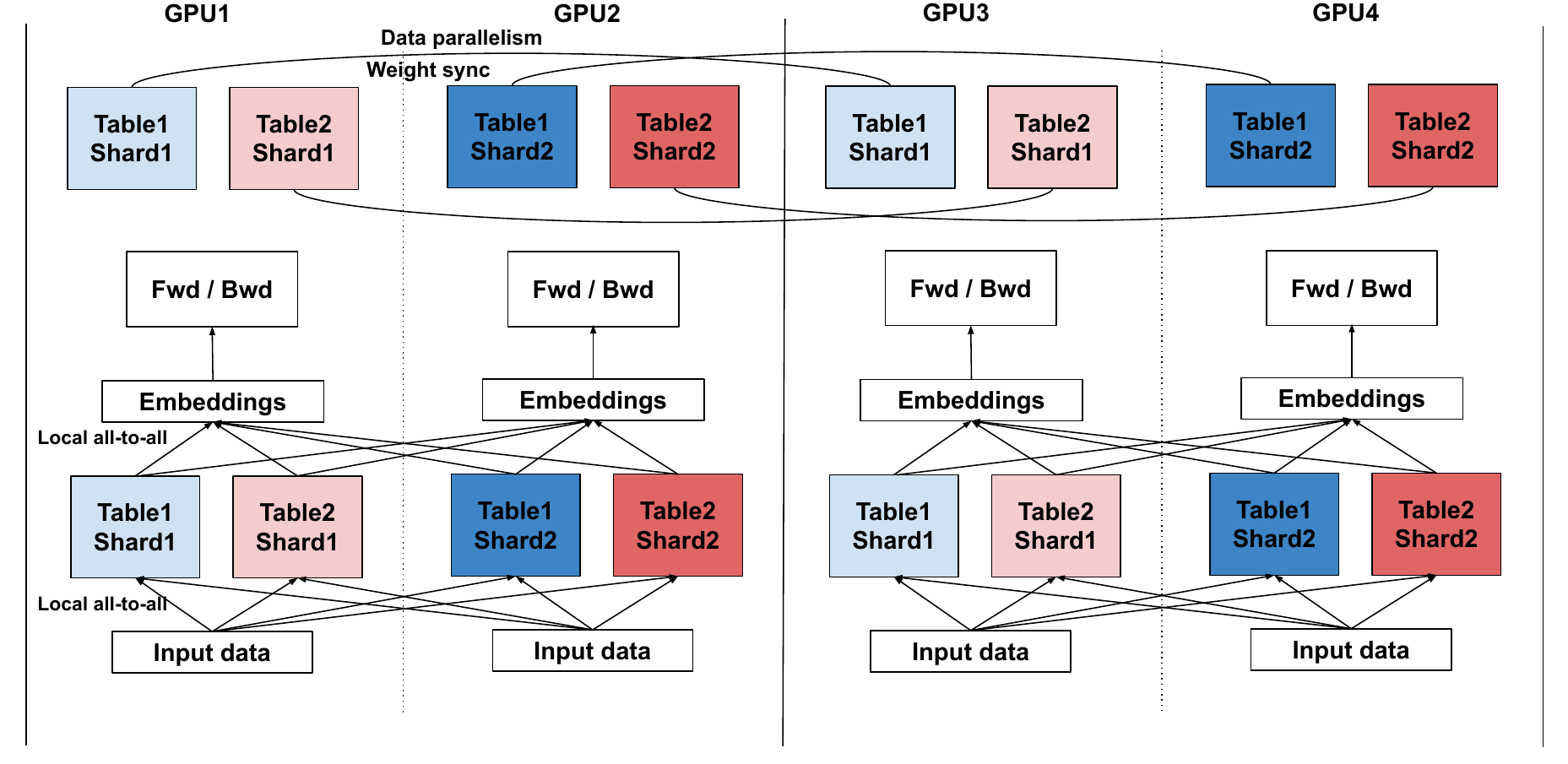}
    \caption{Illustration of 2D sparse parallelism design. The system consists of 4 GPUs dividing into 2 parallelism groups.  The model has two embedding tables, each of which is sharded across 2 GPUs. 
    % In the forward pass, each GPU has its local input data. Each group performs an initial all-to-all to gather the IDs for embedding look-ups, followed by a second all-to-all to distribute the embeddings to the corresponding GPUs. In the backward pass, the system executes a third all-to-all to distribute the gradients corresponding to each embedding. After completing one training loop, an all-reduce communication is conducted to ensure synchronization of the tables across the parallelism groups.
    }
    \label{fig:2D_sys_design}
\end{figure*}

Unlike traditional model parallelism approaches that directly distribute embedding tables across all GPUs, 2D sparse parallelism introduces data parallelism on top of model parallelism. This results in a two-dimensional hierarchical parallelism, which is the basis for the name. 
Specifically, consider a total of $T$ GPUs involved in the training process. We first divide these GPUs into $M$ parallelism groups. Assuming $T$ is divisible by $M$, each parallelism group will consist of $N=T/M$ GPUs. Each parallelism group maintains a complete replica of all the embedding tables, which are fully distributed across the set of GPUs within the parallelism group.
Consequently, there will be a total of $M$ replicas of the embedding tables. Within each replica, model parallelism is employed, while data parallelism is applied across the different replicas.
We illustrate our 2D sparse parallelism strategy in Figure~\ref{fig:2D_sys_design}.
% In the forward pass, each parallelism group performs an initial all-to-all to gather the IDs for embedding look-ups, followed by a second all-to-all to distribute the embeddings to the corresponding GPUs. In the backward pass, the system executes a third all-to-all to distribute the gradients corresponding to each embedding. After completing one training loop, an all-reduce communication is conducted to ensure synchronization of the tables across the parallelism groups.

In our proposed 2D sparse parallelism strategy, embedding tables are distributed across $N$ GPUs within each parallelism group, rather than across the entire set of $T$ GPUs. By distributing the embedding table replicas across a smaller number of GPUs, it becomes easier to achieve balanced sharding, thereby reducing the likelihood of straggler issues.
During the training process, each parallelism group independently performs embedding look-ups for IDs that are present in its respective data batch, with all-to-all communication confined to its set of GPUs. Consequently, it reduces the total number of ID look-ups and accelerates all-to-all communication compared to the traditional full model parallelism methods. Furthermore, since each group's data batch is only a fraction ($i.e. 1/M$) of the global data batch, the memory overhead for embedding look-ups is significantly reduced.
It is important to note that, for memory optimization, the backward and optimization processes are fused for sparse embedding tables \cite{khudia2021fbgemm}. As a result, rather than synchronizing gradients as is common in distributed data parallelism, our 2D sparse parallelism approach synchronizes table weights and optimizer states. We further elaborate on the implications of this synchronization strategy on the development of the training algorithm in Section~\ref{Sec.2D algorithm}.

\subsection{Training Algorithm} \label{Sec.2D algorithm}
% !TEX root = ../main.tex

% \begin{algorithm}
% \caption{Row-wise AdaGrad under 2D sparse parallelism}
% \begin{algorithmic}[1]
% \label{alg: row-wise AdaGrad w./ 2D}
% \FOR{iteration $k=1,\cdots$}
%     \FOR{Rank-$n$ $\in$ $m$-th 2D-group}
%         \FOR{$i$-th sharded row in Rank-$n$}
%             \STATE Back-propagation and within-group All2All for $g_{k,i}^{(m)}$
%             \STATE Update 2nd moment $v_{k-\frac{1}{2},i}^{(m)} = v_{k-1,i}^{(m)} + \|g_{k,i}^{(m)}\|_2^2$
%             \STATE Update local weights $w_{k-\frac{1}{2},i}^{(m)}\!= \!w_{k-1,i}^{(m)} \!-\frac{\eta}{\sqrt{v_{k-\frac{1}{2},i}^{(m)}}+\epsilon} \!\cdot g_{k,i}^{(m)}$
%         \ENDFOR
%     \ENDFOR
% \STATE Weight Sync: $w_{k}^{(j)}\!\! =\!\! \frac{1}{M} \sum_{m\in\text{groups}}w_{k-\frac{1}{2}}^{(m)},$ $\forall j$
% \STATE Moment Sync: $v_{k}^{(j)}\!\! =\!\! \frac{1}{M} \sum_{m\in \text{groups}}v_{k-\frac{1}{2}}^{(m)},$ $\forall j$
% \ENDFOR
% \end{algorithmic}
% \end{algorithm}

The row-wise AdaGrad optimizer is widely recognized as the most effective algorithm for optimizing embedding tables in DLRM training \cite{mudigere2022software,ghaemmaghami2022learning,zhang2024wukong}.
It offers both faster convergence and better generalization compared to the conventional Stochastic Gradient Descent (SGD) optimizer.
In this context, we will concentrate on the row-wise sparse AdaGrad optimizer under the 2D sparse parallelism.
Specifically, the row-wise sparse AdaGrad applies the accumulated row-wise gradient norm to adaptively scale the effective learning rate $\eta_{k,i} = \eta / (\sqrt{v_{k,i}} + \epsilon)$ with $v_{k,i}=\sum_{s=1}^{k} \|g_{s,i}\|_2^2$. Here $g_{s,i}$ is the gradient for the $i$-th row at the $s$-th iteration, $\eta$ is the base learning rate and $\epsilon>0$ is a constant for the numerical stability.
Notably, unlike the original AdaGrad \cite{duchi2011adaptive}, which adapts the learning rate for each individual parameter, the row-wise sparse AdaGrad adapts the learning rate at the row level. This design choice requires only an additional scalar memory cost for each row, making it well-suited for large-scale DLRM with extensive embedding tables.

\begin{algorithm}
\caption{{\color{blue}{Moment-scaled}} row-wise AdaGrad under 2D sparse parallelism}
\begin{algorithmic}[1]
\label{alg: row-wise AdaGrad w./ 2D}
\FOR{iteration $k=1,\cdots$}
    \FOR{GPU-$n$ $\in$ $m$-th 2D-group}
        \FOR{$i$-th sharded row in GPU-$n$}
            \STATE Back-propagation and within-group all-to-all for $g_{k,i}^{(m)}$
            \STATE Update 2nd Moment $v_{k-\frac{1}{2},i}^{(m)} = v_{k-1,i}^{(m)} + \|g_{k,i}^{(m)}\|_2^2$
            \STATE Update local weight $w_{k-\frac{1}{2},i}^{(m)}\!= \!w_{k-1,i}^{(m)} \!-\!\!\frac{\eta}{\sqrt{v_{k-\frac{1}{2},i}^{(m)}{\color{blue}{\mathlarger{\mathbf{/c}}}}}+\epsilon} \!\cdot\! g_{k,i}^{(m)}$ (\color{blue} with scaling factor ${\mathlarger{\mathbf{c}}}$)
        \ENDFOR
    \ENDFOR
\STATE Weight Sync: $w_{k}^{(j)}\!\! =\!\! \frac{1}{M} \sum_{m\in\text{groups}}w_{k-\frac{1}{2}}^{(m)},$ $\forall j$
\STATE Moment Sync: $v_{k}^{(j)}\!\! =\!\! \frac{1}{M} \sum_{m\in \text{groups}}v_{k-\frac{1}{2}}^{(m)},$ $\forall j$
\ENDFOR
\end{algorithmic}
\end{algorithm}

We have Algorithm~\ref{alg: row-wise AdaGrad w./ 2D} (black part) as the integration of the row-wise AdaGrad optimizer under the proposed 2D sparse parallelism.
Suppose table weight replica $w_i^{(m)}$ is sharded on GPU-$n$ in the $m$-th 2D group.
At the $k$-th iteration, GPU-$n$ in the $m$-th 2D group does the local back-propagation to obtain the gradient based on the local data batch $B_{k,m,n}$. Then within each 2D group, the GPUs conduct an all-to-all communication within the group to aggregate group-level gradients for the locally sharded sparse table weights, i.e. $g_{k,i}^{(m)}=\frac{1}{\sum_{n}|B_{k,m,n}|}\sum_{n\in \text{group}_m}\sum_{\zeta \in B_{k,m,n}}\frac{\partial l}{\partial w_i}|_{(\zeta,w_i^{(m)})}$, where $\frac{\partial l}{\partial w_i}|_{(\zeta,w_i^{(m)})}$ represents the gradient of $w_i$ with the loss $l$ evaluated on data sample $\zeta$ and table weight value $w_i^{(m)}$.
With the group-level gradient, each GPU updates the 2nd-order moment and local table weight following the conventional row-wise sparse AdaGrad in parallel. Before ending the $k$-th iteration, the GPUs run the cross-group all-reduce to synchronize the local updated table weights and moments, so that the table weights and optimizer states of each group will keep consensus. 
Note that, if we set the group number as 1 and remove the cross-group all-reduce step, then Algorithm~\ref{alg: row-wise AdaGrad w./ 2D} is exactly the original row-wise sparse AdaGrad without 2D sparse parallelism (i.e. traditional full model parallelism).
However, our preliminary experiments in Figure~\ref{fig:ne_experiments} (a) observes a significant model performance loss of Algorithm~\ref{alg: row-wise AdaGrad w./ 2D}. 
% In what follows, we will focus on the algorithmic analysis to understand the performance loss and then propose the mitigation solution.

After analyzing the discrepancy of the row-wise AdaGrad with and without the 2D sparse parallelism 
% (see Appendix~\ref{sec. proof} in Supplementary Material)
% (see Appendix C in Supplementary Material)
, we have the following proposition to explain the performance loss:

\begin{prop}\label{Prop: vt property}
Under the 2D sparse parallelism, assume that the gradients 
% $\{g_{k,i}^{(m)},~ \forall m\}$
are i.i.d. with respect to the groups. Then it holds that i). the expectation of 2nd-order moments after local updates are the same across the groups, i.e. $\mathbb{E}[v_{k,i}^{(m),\text{2D}}]=\mathbb{E}[v_{k-\frac{1}{2},i}^{(j),\text{2D}}]$, $\forall m,j$; ii). the expected increment of 2nd-order moment under the 2D sparse parallelism is larger than that without the 2D sparse parallelism, i.e. $\mathbb{E}[v_{k,i}^{(m),\text{2D}}-v_{k-1,i}^{(m),\text{2D}}] \ge \mathbb{E}[v_{k,i}^{\text{Non-2D}}-v_{k-1,i}^{\text{Non-2D}}]$, $\forall m$.
\end{prop}
% From Proposition~\ref{Prop: vt property} i), it can approximate the weight updates $w_{k-1,i}^{\text{2D}}- w_{k,i}^{(m),\text{2D}}\approx \frac{\eta}{\sqrt{v_{k,i}^{(m),\text{2D}}} + \epsilon} \times \frac{1}{M}\sum_{m}g_{k,i}^{(m)}$ under the expectation.
From Proposition~\ref{Prop: vt property}, we would conclude that the 2nd-order moment under 2D sparse parallelism $v_{k,i}^{(j),\text{2D}}$ would grow faster than without 2D sparse parallelism $v_{k,i}^{\text{Non-2D}}$. 
Thus, it can be found that the effective learning rate under 2D sparse parallelism would be \textit{smaller} than that without 2D sparse parallelism.  
Such an unintentional learning rate reduction causes the model performance drop.

To fix the learning rate reduction caused by the 2D sparse parallelism, we propose a simple while effective solution, named the \textit{moment-scaled} row-wise sparse AdaGrad (see the blue part in Algorithm~\ref{alg: row-wise AdaGrad w./ 2D}). 
Basically, it introduces a constant factor $c>0$ to scale the 2nd-order moment when computing the effective learning rate (line 6 in Algorithm~\ref{alg: row-wise AdaGrad w./ 2D}): 
Instead of directly using $v_{k-\frac{1}{2},i}^{(m)}$ to adjust the effective learning rate, it scales down the 2nd-order moment and sets the effective learning rate as $\eta/\sqrt{v_{k-\frac{1}{2},i}^{(m)}{{/c}}+\epsilon}$.
With larger value of $c$, the effective learning rate will be larger, so that it can compensate the unexpected learning rate reduction caused by the training paradigm shift.
For the value selection of $c$, we provide the following rule:
\begin{srule}\label{rule: scaling factor}
 With the assumption in Proposition~\ref{Prop: vt property}, the range of $c$ is between $(0, M]$. If the gradients have high noise-signal ratio, then it would prefer large value of $c$.
\end{srule}
\vspace{-.05in}
\noindent 
% Due to limited space, we relegate the detailed derivation and justification in 
% Appendix C in Supplementary Material. 
% Appendix~\ref{sec. proof} in Supplementary Material. 
We also conduct the ablation studies to understand the implication of this hyper-parameter in Section~\ref{Sec.experiments}.
Note that our algorithm can be easily extended to other optimization frameworks which adapt the effective learning rate with 2nd-order moment, such as RMSProp \cite{tieleman2012lecture} and Adam \cite{kingma2014adam}, etc. 
% In our experiments, we conduct the ablation studies to understand the implication of this scaling factor.
% We will find that by appropriately choosing the value of $c$, it can achieve model performance parity under 2D sparse parallelism.

\section{Experiments} \label{Sec.experiments}
% !TEX root = ../main.tex

% In this section, we present the results of our experimental evaluation of the proposed 2D sparse parallelism on real-world recommendation system models, comparing its efficacy with the traditional fully parallelism approach on embedding tables. 
% The experiment specifications are provided in Section~\ref{Sec.experiment setup}. Our experiments are organized into three aspects.
% Section~\ref{Sec.NE experiments} focuses on evaluating the model performance when training with the proposed 2D sparse parallelism and the momentum-scaled row-wise AdaGrad algorithm.
% Section~\ref{Sec.perf experiments} analyzes the system performance of the proposed 2D sparse parallelism, comparing it with the traditional fully parallelism approach.
% Finally, we verify the scaling curve of our 2D sparse parallelism using a massive GPU training system, demonstrating its linear scalability in Section~\ref{Sec.scaling experiments}.

% \subsection{Experimental Setup}\label{Sec.experiment setup}
In this section, we will verify the efficacy of our proposed 2D sparse parallelism as well as the moment-scaled AdaGrad optimizer. 
Due to the space limitation, we will mainly report the experimental results of the industrial real-world recommendation tasks, which have extensive embedding tables.
% while the experiments on public datasets are provided in 
% Appendix~\ref{Appendix:experiment} in Supplementary Material.
% Appendix D in Supplementary Material.
Specifically, our evaluation involves two distinct models: one Click-Through Rate (CTR) model studied in \cite{zhang2022dhen} and an external large foundation model (ExFM) proposed in \cite{liang2025external}. 
The CTR model has a total embedding table size of 0.5TB while ExFM is 1.7TB.

\subsection{Model Performance}\label{Sec.NE experiments}
We first evaluate the model performance when training with the proposed approach.
We adopt the normalized entropy (NE) \cite{he2014practical} as the metric to measure the model performance. 
The baseline is the traditional full model parallelism approach with row-wise AdaGrad.
We report the NE gap for the comparison, where a larger gap means more significant model performance loss,
a NE difference of 0.02\% is considered as significant.

\begin{figure*}
    \centering
    \begin{tabular}{cc}
        \includegraphics[width=0.43\textwidth]{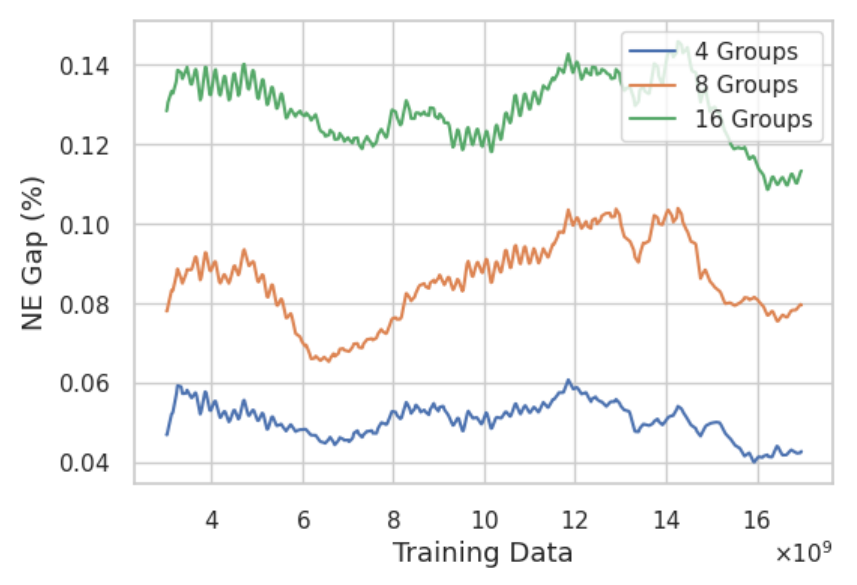} & 
        \includegraphics[width=0.415\textwidth]{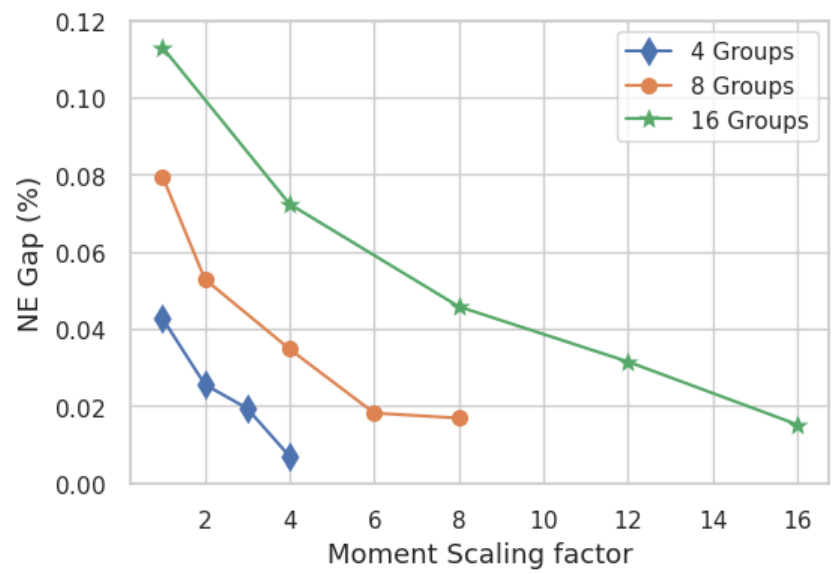} \\
        (a) & (b) 
    \end{tabular}
    \caption{Normalized entropy (NE) gaps on the CTR model, where $0.02\%$ can be considered as the significant threshold. (a) is NE gap of directly applying 2D sparse parallelism; (b) NE gap vs moment scaling factor after 15 billion data training. The model performance loss caused by 2D sparse parallelism can be mitigated with moment-scale row-wise AdaGrad. }
    \label{fig:ne_experiments}
    \vspace{-.2in}
\end{figure*}

% \begin{figure}
%     \centering
%     \includegraphics[width=0.98\linewidth]{Figures/CTR_group_ne_gap.pdf}
%     \vspace{-.1in}
%     \caption{Normalized entropy (NE) gaps of directly applying 2D sparse parallelism on the downstream CTR model. }
%     \label{fig:ctr_2d_ne_gap}
% \end{figure}

We start the studies with the CTR model. 
We trained the model with 256 GPUs and set the batch size as 4096 for each GPU.
For 2D sparse parallelism, we vary the group number in $\{4,8,16\}$.
In Figure~\ref{fig:ne_experiments} (a), we show the NE gap when directly apply 2D sparse parallelism without any algorithm change.
It can be seen that the NE gaps for all the three group numbers are significantly above the 0.02\% variance threshold after training on 15 billion samples.
The NE gap is larger with more groups. This matches with our analysis that a larger group number results in a smaller effective learning rate, which causes insufficient model training.
Now we apply the moment-scaled row-wise AdaGrad as the mitigation. With different group numbers, we tune the scaling factor for an ablation study.
The results are shown in Figure~\ref{fig:ne_experiments} (b). 
It can be seen that as the scaling factor approaches the group number, the NE gap is tending to close (<0.02\%).
% From our analysis, it means the model has large noise in the gradients. 
% Thus, for our in-house models, we recommend to use the group number as the scaling factor.
% \begin{figure}
%     \centering
%     \includegraphics[width=0.95\linewidth]{Figures/ctr_ne_scaling.pdf}
%     \vspace{-.1in}
%     \caption{Normalized entropy (NE) gap vs moment scaling factor on the downstream CTR model. The NE gap is at about 15 billion data training.}
%     \label{fig:ctr_ne_scaling}
% \end{figure}

% \begin{figure}
%     \centering
%     \includegraphics[width=0.98\linewidth]{Figures/omnifm_ne.pdf}
%     \vspace{-.1in}
%     \caption{Normalized entropy (NE) gaps of 2D sparse parallelism on the exLarge foundation model. The model is trained with 1024 GPUs and 2D sparse parallelism adopts 4 groups. }
%     \label{fig:omnifm_ne}
% \end{figure}

\begin{figure}
   \includegraphics[width=\linewidth]{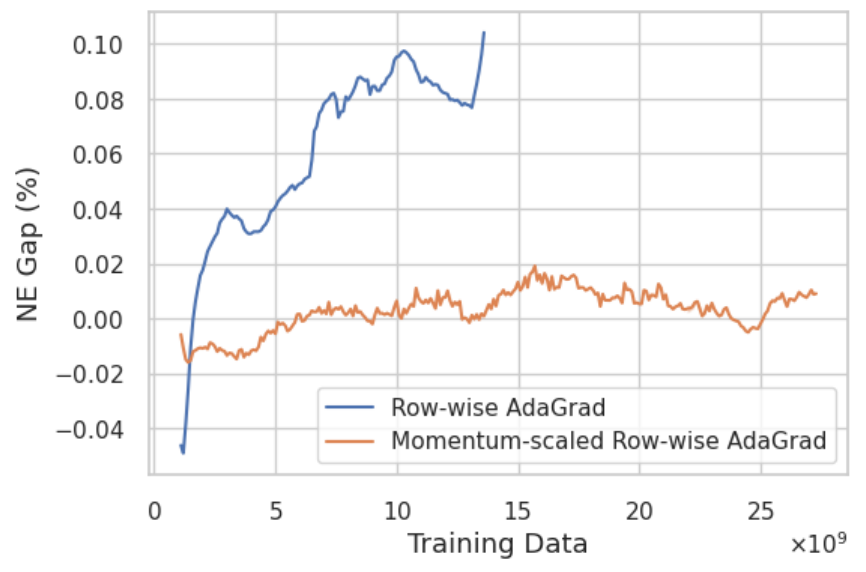}
    \caption{Normalized entropy (NE) gaps of 2D sparse parallelism on the ExFM with 1024 GPUs. }
    \label{fig:omnifm_ne}
    \vspace{-.2in}
\end{figure}
Then we extend to the ExFM.
We train the model with 1024 GPUs and use 256 GPUs per group, 
thus total 4 groups. Based on the learning from the CTR model, here we set the moment scaling factor as 4. 
From Figure~\ref{fig:omnifm_ne}, it observes about 0.1\% NE regression when adopting 2D sparse parallelism with the original row-wise AdaGrad.
When training with the proposed moment-scaled row-wise AdaGrad, the NE gap is fully closed with <0.02\% variance. Thus, we conclude that, combining the moment-scaled row-wise Adagrad with a proper scaling factor, the 2D sparse parallelism has no concern on the model performance. 
Additionally, we recommend to use the group number as the moment scaling factor to simplify the algorithm tuning.

\subsection{System Performance}\label{Sec.perf experiments}
Now we evaluate the training efficiency of our proposed 2D sparse parallelism approach by varying the number of parallelism groups and comparing its performance to traditional full model parallelism.
We consider three key metrics: training queries per second (QPS), GPU memory usage percentage, and embedding look-up imbalance ratio. 
The QPS is measuring the number of data processed per second, it is the metric for training speed.
The imbalance ratio is defined as the value of the maximum embedding look-up latency over the average look-up latency across GPUs.
The larger imbalance ratio means a more severe straggler issue.
The CTR model is trained with 256 GPUs and 4096 batch size per GPU; while the ExFM is trained with 1024 GPUs and 896 batch size per GPU.
% We also report different kernel costs in Figure~\ref{fig:sys_perf_breakdown} for better understanding of these training efficiency metrics.  

\begin{table*}
\centering
\caption{Efficiency comparison on QPS, memory and imbalance ratio (imb. ratio).}
\vspace{.05in}
\label{table: system_perf}
% \begin{tabular}{c|c|c|c|c|c|c}
% \hline
%  \multirow{2}{*}{Parallelism Strategy}   & \multicolumn{3}{c|}{Downstream CTR model} & \multicolumn{3}{c}{ExLarge Foundation Model} \\ 
%  \cline{2-7}
%   & QPS & Peak/Avg. Memory  &Imb. Ratio  & QPS &  Peak/Avg. Memory  & Imb. Ratio   \\
%   \hline
%   Full Model Parallelism  &  $1.53\times 10^6$   & 72.40 \% / 49.66 \% & 5.70 &  $5.51\times 10^5$   & 99.44 \% / 76.95 \% & 5.18 \\
%   2D Sparse Parallelism, 2 Groups &  $2.34\times 10^6$  & 53.65 \% / 47.65 \% & 2.01  &  $6.31\times 10^5$  & 85.57\% / 79.15 \% & 2.23 \\

%   2D Sparse Parallelism, 4 Groups  &  $2.66\times 10^6$  & 54.74 \% / 50.59 \% & 1.57    &  $6.54\times 10^5$  & 87.77 \% / 82.81 \% & 1.65 \\

%   2D Sparse Parallelism, 8 Groups  &  $2.26\times 10^6$  & 69.90 \% / 62.91 \% & 1.53   &  $6.32\times 10^5$  & 94.85 \% / 90.22 \% & 1.63 \\
% \hline
% \end{tabular}
{ \begin{tabular}{c|c|c|c|c}
\hline
 % \cline{2-7}
  Model & Parallelism Strategy & QPS & Peak/Avg. Memory  &Imb. Ratio    \\
  \hline 
 \multirow{4}{*}{CTR Model}  & Model Parallelism  &  $1.53\times 10^6$   & 72.40 \% / 49.66 \% & 5.70 \\
  
       &  2D Parallelism, 2 Groups &  $2.34\times 10^6$  & 53.65 \% / 47.65 \% & 2.01   \\

  & 2D Parallelism, 4 Groups  &  $2.66\times 10^6$  & 54.74 \% / 50.59 \% & 1.57 \\

 & 2D Parallelism, 8 Groups  &  $2.26\times 10^6$  & 69.90 \% / 62.91 \% & 1.53   \\
\hline
 \multirow{4}{*}{ExFM} &  Model Parallelism   &  $5.51\times 10^5$   & 99.44 \% / 76.95 \% & 5.18  \\

  &  2D Parallelism, 2 Groups &  $6.31\times 10^5$  & 85.57\% / 79.15 \% & 2.23 \\

 & 2D Parallelism, 4 Groups &   $6.54\times 10^5$  & 87.77 \% / 82.81 \% & 1.65 \\
  & 2D Parallelism, 8 Groups&  $6.32\times 10^5$  & 94.85 \% / 90.22 \% & 1.63 \\
\hline
\end{tabular}}
\end{table*}

The experimental results are summarized in Table~\ref{table: system_perf}.
% which confirms the training efficacy of our 2D sparse parallelism approach.
Notably, the traditional model parallelism approach exhibits an imbalance ratio above $5$ for the two models.
This indicates that the slowest GPU takes more than 5 times longer to complete the embedding look-up compared to the average.
In contrast, our approach substantially reduces the imbalance ratio. With 4 parallelism groups, the imbalance ratio drops below $2$, effectively eliminating stragglers and promoting a more balanced system. Furthermore, the memory usage analysis reveals that adopting our 2D sparse parallelism leads to a substantial reduction in peak memory compared to model parallelism, with savings of approximately 20\% on the downstream CTR model and about 10\% on the ExFM.
Moreover, our approach yields significant improvements in training speed. Specifically, with 4 parallelism groups, the training QPS is nearly doubled on the CTR model, and improved by about 20\% on the ExFM. These findings collectively demonstrate the effectiveness of our 2D sparse parallelism approach in achieving better system balance, reducing memory requirements, and accelerating training speeds.
\begin{figure*}
    \centering
    \includegraphics[width=0.8\linewidth]{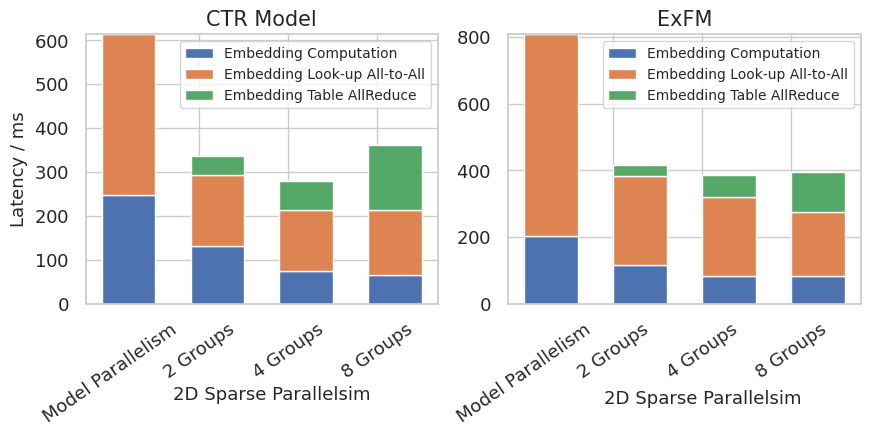}
    \vspace{-.1in}
    \caption{Maximum kernel costs under different parallelism strategies.}
    \label{fig:sys_perf_breakdown}
\end{figure*}

While our experiments demonstrate the benefits of 2D sparse parallelism, they also reveal that increasing the number of parallelism groups does not always lead to improved training efficiency. As shown in Table~\ref{table: system_perf}, the optimal memory and QPS are achieved when the number of parallelism groups is 4 for both models, whereas using 8 groups results in degraded performance.
A closer examination of the memory consumption reveals that when the number of parallelism groups increases from 4 to 8, the number of GPUs in each group is halved, requiring each GPU to host twice the size of embedding tables. This leads to increased memory consumption.
Furthermore, the training QPS decreases by 15\% on the CTR model and 5\% on the ExFM. To understand the underlying causes of this decline, we analyzed the kernel costs in Figure~\ref{fig:sys_perf_breakdown}. While the latency of computation and embedding look-up all-to-all decreases with more groups, the table all-reduce latency increases.
For example, on the CTR model, the total latency rises from approximately 280ms with 4 groups to around 350ms with 8 groups. The primary contributor to this increase is the table all-reduce synchronization latency, which negatively impacts the training QPS.
These findings highlight the importance of selecting an optimal number of parallelism groups when applying 2D sparse parallelism in practice. 

\subsection{Massive GPU Scaling}\label{Sec.scaling experiments}
We conducted experiments to validate the scalability of our proposed 2D sparse parallelism, focusing on the ExFM with massive GPU training. 
We set the batch size at 1152 for each GPU. For our 2D sparse parallelism, we fix 256 GPUs per group, and this is equivalent to the traditional model parallelism approach when training with 256 GPUs.
We summarize the QPS and memory results in Table~\ref{table: QPS_scaling}.
Here we introduce a new metric named QPS scaling factor to measure the scaling efficiency. 
It is defined as the ratio of the QPS speed-up over the multiplier of GPU number increasing.
From Table~\ref{table: QPS_scaling}, it is evident that, with the traditional model parallelism approach, QPS increases sub-linearly as the number of GPUs scales up. 
When training with 1024 GPUs, the QPS scaling factor deteriorates to less than 80\%.
Also, traditional model parallleism encountered out-of-memory failures when training with more than 1024 GPUs.
This makes further GPU scaling infeasible with the traditional approach.
In contrast, our proposed 2D sparse parallelism does not encounter memory bottlenecks during GPU scaling.
\begin{figure}
   \includegraphics[width=.8\linewidth]{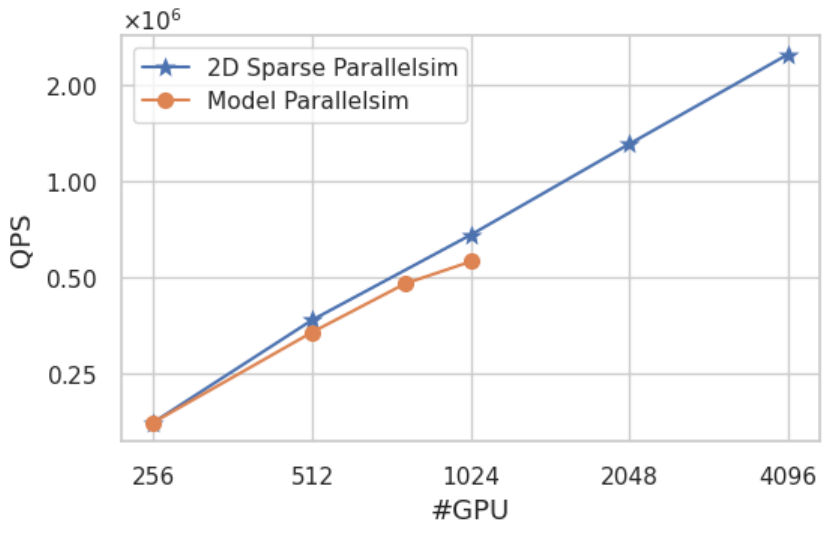}
    \caption{QPS vs \#GPU on the ExFM. The traditional full parallelism approach hits out-of-memory failure when training beyond 1024 GPUs. The proposed 2D sparse parallelism can achieve nearly linear scalability on QPS.}
    \label{fig:omnifm_gpu_scaling}
\end{figure}
With 256 GPUs per group, the memory cost of embedding table and look-up would maintain at a fixed scale even as total GPU number increases, so does the peak GPU memory usage (about 95\%).
Compared to the traditional model parallelism approach, our 2D sparse parallelism achieves a 9.1\% and 21.2\% improvement in QPS when training with 512 GPUs and 1024 GPUs, respectively.
For further GPU scaling up, the QPS scaling factor remains 95\% up to 2048 GPUs. 
The scaling factor is 90\% when training with 4096 GPUs. 
The slight regression is because cross-building communication is triggered. The long-distance communication introduces additional latency, hindering the scaling factor. 
For better visualization, we show the QPS scaling in Figure~\ref{fig:omnifm_gpu_scaling}, where our proposed 2D sparse parallelism achieves nearly linear scalability, implying the GPU power is fully leveraged.
% \begin{figure}
%   \begin{center}
%     \includegraphics[width=\linewidth]{Figures/omnifm_scaling.pdf}
%   \end{center}
%   \caption{QPS vs \#GPU on the exLarge foundation model. The traditional fully parallelism approach hits out-of-memory failure when training beyond 1024 GPUs.}
%     \label{fig:omnifm_gpu_scaling}
% \end{figure}

% \begin{table}
% \centering
% \caption{GPU scaling on the exLarge foundation model. The batch size is 1152 per GPU. Speed-up and scaling factor are calculated based on results of 256 GPUs. The full model parallelism approach hits out-of-memory (OOM) failure with \#GPU > 1024.}
% \label{table: QPS_scaling}
% \begin{tabular}{c|c|c|c|c|c|c|c|c}
% \hline
%  \multirow{2}{*}{\# GPU}   & \multicolumn{4}{c|}{Full Model Parallelism Approach} & \multicolumn{4}{c}{2D Sparse Parallelism} \\ 
%  \cline{2-9}
%  & Peak Memory & QPS & Speed-up & Scaling Factor & Peak Memory & QPS & Speed-up & Scaling Factor\\
%   \hline
%   256  &  96.54\% & $1.76\times 10^5$ & - & - & 96.54\% & $1.76\times 10^5$ & - & - \\
%   512  & 95.47\% & $3.37\times 10^5$ & 1.91x & 95\% &96.59\% & $3.56\times 10^5$ & 2.02x & 100\% \\
%   1024 & 100\% & $5.61\times 10^5$ & 3.19x & 79.7\% &94.95\% & $6.76\times 10^5$ & 3.84x & 96\% \\
%   2048 & OOM & N/A &N/A & N/A&95.07\% & $1.34\times 10^6$ & 7.61x & 95\%\\
%   4096 & OOM &N/A &N/A &N/A &95.56\% & $2.53\times 10^6$ & 14.38x & 90\% \\
% \hline
% \end{tabular}
% \end{table}

\begin{table*}
\centering
\caption{GPU scaling on the ExFM. The batch size is fixed as 1152. The full model parallelism approach hits out-of-memory (OOM) failure with \#GPU > 1024.}
\vspace{.05in}
\label{table: QPS_scaling}
\begin{tabular}{c|c|c|c|c|c|c}
\hline
 \multirow{2}{*}{\# GPU}   & \multicolumn{3}{c|}{Full Model Parallelism Approach} & \multicolumn{3}{c}{2D Sparse Parallelism} \\ 
 \cline{2-7}
 & Memory & QPS & Scaling Factor & Memory & QPS & Scaling Factor\\
  \hline
  256  &  96.54\% & $1.76\times 10^5$  & - & 96.54\% & $1.76\times 10^5$ & - \\
  512  & 95.47\% & $3.37\times 10^5$  & 95\% &96.59\% & $3.56\times 10^5$ &  100\% \\
  1024 & 100\% & $5.61\times 10^5$  & 79.7\% &94.95\% & $6.76\times 10^5$ &  96\% \\
  2048 & OOM & N/A  & N/A&95.07\% & $1.34\times 10^6$  & 95\%\\
  4096 & OOM &N/A &N/A  &95.56\% & $2.53\times 10^6$  & 90\% \\
\hline
\end{tabular}
\end{table*}

% before applying 2D  QPS

% after applying 2D QPS

% scaling plot num of trainers vs  QPS

% compute the scaling factor before and after applying 2D

\section{Deployment Considerations and Strategies} \label{Sec.2D optim}
% !TEX root = ../main.tex

While our proposed 2D sparse parallelism enhances the massive GPU efficiency of large-scale DLRMs, we admit two overheads that cannot be neglected: (1) memory overhead resulting from table replication, and (2) communication latency due to the synchronization of table replicas and optimizer states.
In the followings, we will focus on the two overheads and provide corresponding mitigation solutions. 

\paragraph{Table Replication Memory Overhead.} Assume the size of all the sparse embedding tables is $S$ GB. Compared with the traditional full model parallelism approach, the 2D sparse parallelism has the table replication memory overhead for each GPU as $\mathcal{M}_{\text{overhead}} = S(M-1)/T. \label{Eq: Memory overhead}$
Since the memory overhead is directly proportional to the number of groups $M$ and inversely proportional to the total number of training GPUs $T$.
Therefore, it becomes less of a concern when training with a large number of GPUs and utilizing a limited number of 2D sharding groups.
To further reduce the memory overhead, we can reduce the size of the sparse embedding tables $S$ by leveraging the advanced compression and quantization techniques, including the FP8 quantization \cite{kuzmin2022fp8}, unified embedding \cite{coleman2023unified} and tensor-train compression \cite{yin2021tt} etc.

\paragraph{Replica Synchronization Latency.} Compared with the traditional full model parallelism approach, our 2D sparse parallelism introduces an additional synchronization communication for the table replica and optimizer states. By adopting the ring all-reduce synchronization, the communication latency per rank can be estimated as 
\begin{align}
    \mathcal{L}_{\text{sync}} = 2 S(M-1)/(T\cdot B_{\text{sync}}) = 2 \mathcal{M}_{\text{overhead}}/B_{\text{sync}}, \label{Eq: sync latency}
\end{align}
where $B_{\text{sync}}$ is the the communication bandwidth to execute the synchronization.
Apparently, as the communication latency $\mathcal{L}_{\text{sync}}$ is proportional to the memory overhead $\mathcal{M}_{\text{overhead}}$. The aforementioned strategies for  $\mathcal{M}_{\text{overhead}}$ reduction can also help reduce the communication latency.
To further reduce latency, we can leverage intra-host communication to enhance $B_{\text{sync}}$. For instance, consider the setup with 8 GPUs connected to the same CPU host. Within this setup, GPUs on the same CPU host can maintain replicas of identical segments of sparse embedding tables. When the group number are chose from $\{2,4,8\}$, synchronization will be confined to the same host. Intra-host communication is markedly faster than inter-host communication, yielding a 7x improvement according to our empirical observations.
Moreover, we can decrease the synchronization frequency and enable local updates, akin to the local SGD approach \cite{stich2019local,haddadpour2019local} used in federated learning paradigms. However, it is important to note that significantly lowering the synchronization frequency may affect training performance. Consequently, careful tuning is required for each specific use case.

\section{Conclusion} \label{Sec.conclusion}

% !TEX root = ../main.tex

% \textit{Dynamic 2D sparse parallelism.} In the above system, we apply the same data parallelism to all embedding tables. However this still might not be optimal from the training efficiency perspective. In real world DLRMs, the sharding imbalance is usually largely caused by some \textit{hot} embedding tables. For these \textit{hot} tables,  we need to look up a significant larger number of rows from these tables in a single training sample compared to ordinary tables. Some typical hot tables are event based feature embedding tables, where a long sequence of event IDs needs to be looked up for a training sample. In this case, one can apply 2D sparse parallelism for these hot tables, which largely mitigate the sharding imbalance with minimal table replication overhead and parameter synchronization latency.

In this paper, we introduced two-dimensional sparse parallelism for embedding tables in large-scale distributed DLRM systems.
Unlike the traditional full model parallelism approach, our strategy organizes GPUs into groups, each maintaining a replica of the embedding table, which allows sparse all-to-all communication to occur within each group, effectively addressing workload imbalance, intensive look-up communication, and excessive memory consumption challenges.
After each training loop, an all-reduce synchronization ensures table weight consensus across replicas. 
To mitigate performance losses due to the shift in training paradigms, we developed a new optimization algorithm called moment-scaled row-wise AdaGrad. By appropriately scaling the moment, this algorithm maintains model performance parity without compromising training efficiency.
Our evaluation on both small-scale CTR model and large-scale foundation model demonstrates significant improvements in training efficiency. It verified the near-linear scalability on the foundation model using up to $4K$ GPUs, establishing a new state-of-the-art benchmark for recommendation model training.

\bibliographystyle{ACM-Reference-Format}
\bibliography{reference}


\begin{thebibliography}{38}

%%% the Web address.

\ifx \showCODEN    \undefined \def \showCODEN     #1{\unskip}     \fi
\ifx \showISBNx    \undefined \def \showISBNx     #1{\unskip}     \fi
\ifx \showISBNxiii \undefined \def \showISBNxiii  #1{\unskip}     \fi
\ifx \showISSN     \undefined \def \showISSN      #1{\unskip}     \fi
\ifx \showLCCN     \undefined \def \showLCCN      #1{\unskip}     \fi
\ifx \shownote     \undefined \def \shownote      #1{#1}          \fi
\ifx \showarticletitle \undefined \def \showarticletitle #1{#1}   \fi
\ifx \showURL      \undefined \def \showURL       {\relax}        \fi
% invisible to TeX
\providecommand\bibfield[2]{#2}
\providecommand\bibinfo[2]{#2}
\providecommand\natexlab[1]{#1}
\providecommand\showeprint[2][]{arXiv:#2}

\bibitem[Ardalani et~al\mbox{.}(2022)]%
        {ardalani2022understanding}
\bibfield{author}{\bibinfo{person}{Newsha Ardalani}, \bibinfo{person}{Carole-Jean Wu}, \bibinfo{person}{Zeliang Chen}, \bibinfo{person}{Bhargav Bhushanam}, {and} \bibinfo{person}{Adnan Aziz}.} \bibinfo{year}{2022}\natexlab{}.
\newblock \showarticletitle{Understanding scaling laws for recommendation models}.
\newblock \bibinfo{journal}{\emph{arXiv preprint arXiv:2208.08489}} (\bibinfo{year}{2022}).
\newblock


\bibitem[Cai et~al\mbox{.}(2024)]%
        {cai2024distributed}
\bibfield{author}{\bibinfo{person}{Qiqi Cai}, \bibinfo{person}{Jian Cao}, \bibinfo{person}{Guandong Xu}, {and} \bibinfo{person}{Nengjun Zhu}.} \bibinfo{year}{2024}\natexlab{}.
\newblock \showarticletitle{Distributed Recommendation Systems: Survey and Research Directions}.
\newblock \bibinfo{journal}{\emph{ACM Transactions on Information Systems}} \bibinfo{volume}{43}, \bibinfo{number}{1} (\bibinfo{year}{2024}), \bibinfo{pages}{1--38}.
\newblock


\bibitem[Cheng et~al\mbox{.}(2016)]%
        {cheng2016wide}
\bibfield{author}{\bibinfo{person}{Heng-Tze Cheng}, \bibinfo{person}{Levent Koc}, \bibinfo{person}{Jeremiah Harmsen}, \bibinfo{person}{Tal Shaked}, \bibinfo{person}{Tushar Chandra}, \bibinfo{person}{Hrishi Aradhye}, \bibinfo{person}{Glen Anderson}, \bibinfo{person}{Greg Corrado}, \bibinfo{person}{Wei Chai}, \bibinfo{person}{Mustafa Ispir}, {et~al\mbox{.}}} \bibinfo{year}{2016}\natexlab{}.
\newblock \showarticletitle{Wide \& deep learning for recommender systems}. In \bibinfo{booktitle}{\emph{Proceedings of the 1st workshop on deep learning for recommender systems}}. \bibinfo{pages}{7--10}.
\newblock


\bibitem[Coleman et~al\mbox{.}(2023)]%
        {coleman2023unified}
\bibfield{author}{\bibinfo{person}{Benjamin Coleman}, \bibinfo{person}{Wang-Cheng Kang}, \bibinfo{person}{Matthew Fahrbach}, \bibinfo{person}{Ruoxi Wang}, \bibinfo{person}{Lichan Hong}, \bibinfo{person}{Ed Chi}, {and} \bibinfo{person}{Derek Cheng}.} \bibinfo{year}{2023}\natexlab{}.
\newblock \showarticletitle{Unified Embedding: Battle-tested feature representations for web-scale ML systems}.
\newblock \bibinfo{journal}{\emph{Advances in Neural Information Processing Systems}}  \bibinfo{volume}{36} (\bibinfo{year}{2023}), \bibinfo{pages}{56234--56255}.
\newblock


\bibitem[Covington et~al\mbox{.}(2016)]%
        {covington2016deep}
\bibfield{author}{\bibinfo{person}{Paul Covington}, \bibinfo{person}{Jay Adams}, {and} \bibinfo{person}{Emre Sargin}.} \bibinfo{year}{2016}\natexlab{}.
\newblock \showarticletitle{Deep neural networks for youtube recommendations}. In \bibinfo{booktitle}{\emph{Proceedings of the 10th ACM conference on recommender systems}}. \bibinfo{pages}{191--198}.
\newblock


\bibitem[Duchi et~al\mbox{.}(2011)]%
        {duchi2011adaptive}
\bibfield{author}{\bibinfo{person}{John Duchi}, \bibinfo{person}{Elad Hazan}, {and} \bibinfo{person}{Yoram Singer}.} \bibinfo{year}{2011}\natexlab{}.
\newblock \showarticletitle{Adaptive subgradient methods for online learning and stochastic optimization.}
\newblock \bibinfo{journal}{\emph{Journal of machine learning research}} \bibinfo{volume}{12}, \bibinfo{number}{7} (\bibinfo{year}{2011}).
\newblock


\bibitem[Elkahky et~al\mbox{.}(2015)]%
        {elkahky2015multi}
\bibfield{author}{\bibinfo{person}{Ali~Mamdouh Elkahky}, \bibinfo{person}{Yang Song}, {and} \bibinfo{person}{Xiaodong He}.} \bibinfo{year}{2015}\natexlab{}.
\newblock \showarticletitle{A multi-view deep learning approach for cross domain user modeling in recommendation systems}. In \bibinfo{booktitle}{\emph{Proceedings of the 24th international conference on world wide web}}. \bibinfo{pages}{278--288}.
\newblock


\bibitem[Ghaemmaghami et~al\mbox{.}(2022)]%
        {ghaemmaghami2022learning}
\bibfield{author}{\bibinfo{person}{Benjamin Ghaemmaghami}, \bibinfo{person}{Mustafa Ozdal}, \bibinfo{person}{Rakesh Komuravelli}, \bibinfo{person}{Dmitriy Korchev}, \bibinfo{person}{Dheevatsa Mudigere}, \bibinfo{person}{Krishnakumar Nair}, {and} \bibinfo{person}{Maxim Naumov}.} \bibinfo{year}{2022}\natexlab{}.
\newblock \showarticletitle{Learning to collide: Recommendation system model compression with learned hash functions}.
\newblock \bibinfo{journal}{\emph{arXiv preprint arXiv:2203.15837}} (\bibinfo{year}{2022}).
\newblock


\bibitem[Haddadpour et~al\mbox{.}(2019)]%
        {haddadpour2019local}
\bibfield{author}{\bibinfo{person}{Farzin Haddadpour}, \bibinfo{person}{Mohammad~Mahdi Kamani}, \bibinfo{person}{Mehrdad Mahdavi}, {and} \bibinfo{person}{Viveck Cadambe}.} \bibinfo{year}{2019}\natexlab{}.
\newblock \showarticletitle{Local SGD with periodic averaging: Tighter analysis and adaptive synchronization}.
\newblock \bibinfo{journal}{\emph{Advances in Neural Information Processing Systems}}  \bibinfo{volume}{32} (\bibinfo{year}{2019}).
\newblock


\bibitem[He et~al\mbox{.}(2014)]%
        {he2014practical}
\bibfield{author}{\bibinfo{person}{Xinran He}, \bibinfo{person}{Junfeng Pan}, \bibinfo{person}{Ou Jin}, \bibinfo{person}{Tianbing Xu}, \bibinfo{person}{Bo Liu}, \bibinfo{person}{Tao Xu}, \bibinfo{person}{Yanxin Shi}, \bibinfo{person}{Antoine Atallah}, \bibinfo{person}{Ralf Herbrich}, \bibinfo{person}{Stuart Bowers}, {et~al\mbox{.}}} \bibinfo{year}{2014}\natexlab{}.
\newblock \showarticletitle{Practical lessons from predicting clicks on ads at facebook}. In \bibinfo{booktitle}{\emph{Proceedings of the eighth international workshop on data mining for online advertising}}. \bibinfo{pages}{1--9}.
\newblock


\bibitem[Jha et~al\mbox{.}(2024)]%
        {jha2024mem}
\bibfield{author}{\bibinfo{person}{Gopi~Krishna Jha}, \bibinfo{person}{Anthony Thomas}, \bibinfo{person}{Nilesh Jain}, \bibinfo{person}{Sameh Gobriel}, \bibinfo{person}{Tajana Rosing}, {and} \bibinfo{person}{Ravi Iyer}.} \bibinfo{year}{2024}\natexlab{}.
\newblock \showarticletitle{Mem-rec: Memory efficient recommendation system using alternative representation}. In \bibinfo{booktitle}{\emph{Asian Conference on Machine Learning}}. PMLR, \bibinfo{pages}{518--533}.
\newblock


\bibitem[Kang et~al\mbox{.}(2021)]%
        {kang2021learning}
\bibfield{author}{\bibinfo{person}{Wang-Cheng Kang}, \bibinfo{person}{Derek~Zhiyuan Cheng}, \bibinfo{person}{Tiansheng Yao}, \bibinfo{person}{Xinyang Yi}, \bibinfo{person}{Ting Chen}, \bibinfo{person}{Lichan Hong}, {and} \bibinfo{person}{Ed~H Chi}.} \bibinfo{year}{2021}\natexlab{}.
\newblock \showarticletitle{Learning to embed categorical features without embedding tables for recommendation}. In \bibinfo{booktitle}{\emph{Proceedings of the 27th ACM SIGKDD Conference on Knowledge Discovery \& Data Mining}}. \bibinfo{pages}{840--850}.
\newblock


\bibitem[Khudia et~al\mbox{.}(2021)]%
        {khudia2021fbgemm}
\bibfield{author}{\bibinfo{person}{Daya Khudia}, \bibinfo{person}{Jianyu Huang}, \bibinfo{person}{Protonu Basu}, \bibinfo{person}{Summer Deng}, \bibinfo{person}{Haixin Liu}, \bibinfo{person}{Jongsoo Park}, {and} \bibinfo{person}{Mikhail Smelyanskiy}.} \bibinfo{year}{2021}\natexlab{}.
\newblock \showarticletitle{Fbgemm: Enabling high-performance low-precision deep learning inference}.
\newblock \bibinfo{journal}{\emph{arXiv preprint arXiv:2101.05615}} (\bibinfo{year}{2021}).
\newblock


\bibitem[Kingma and Ba(2014)]%
        {kingma2014adam}
\bibfield{author}{\bibinfo{person}{Diederik~P Kingma} {and} \bibinfo{person}{Jimmy Ba}.} \bibinfo{year}{2014}\natexlab{}.
\newblock \showarticletitle{Adam: A method for stochastic optimization}.
\newblock \bibinfo{journal}{\emph{arXiv preprint arXiv:1412.6980}} (\bibinfo{year}{2014}).
\newblock


\bibitem[Kuzmin et~al\mbox{.}(2022)]%
        {kuzmin2022fp8}
\bibfield{author}{\bibinfo{person}{Andrey Kuzmin}, \bibinfo{person}{Mart Van~Baalen}, \bibinfo{person}{Yuwei Ren}, \bibinfo{person}{Markus Nagel}, \bibinfo{person}{Jorn Peters}, {and} \bibinfo{person}{Tijmen Blankevoort}.} \bibinfo{year}{2022}\natexlab{}.
\newblock \showarticletitle{Fp8 quantization: The power of the exponent}.
\newblock \bibinfo{journal}{\emph{Advances in Neural Information Processing Systems}}  \bibinfo{volume}{35} (\bibinfo{year}{2022}), \bibinfo{pages}{14651--14662}.
\newblock


\bibitem[Liang et~al\mbox{.}(2025)]%
        {liang2025external}
\bibfield{author}{\bibinfo{person}{Mingfu Liang}, \bibinfo{person}{Xi Liu}, \bibinfo{person}{Rong Jin}, \bibinfo{person}{Boyang Liu}, \bibinfo{person}{Qiuling Suo}, \bibinfo{person}{Qinghai Zhou}, \bibinfo{person}{Song Zhou}, \bibinfo{person}{Laming Chen}, \bibinfo{person}{Hua Zheng}, \bibinfo{person}{Zhiyuan Li}, {et~al\mbox{.}}} \bibinfo{year}{2025}\natexlab{}.
\newblock \showarticletitle{External Large Foundation Model: How to Efficiently Serve Trillions of Parameters for Online Ads Recommendation}.
\newblock \bibinfo{journal}{\emph{arXiv preprint arXiv:2502.17494}} (\bibinfo{year}{2025}).
\newblock


\bibitem[Liu et~al\mbox{.}(2024)]%
        {liu2024embedding}
\bibfield{author}{\bibinfo{person}{Shijie Liu}, \bibinfo{person}{Nan Zheng}, \bibinfo{person}{Hui Kang}, \bibinfo{person}{Xavier Simmons}, \bibinfo{person}{Junjie Zhang}, \bibinfo{person}{Matthias Langer}, \bibinfo{person}{Wenjing Zhu}, \bibinfo{person}{Minseok Lee}, {and} \bibinfo{person}{Zehuan Wang}.} \bibinfo{year}{2024}\natexlab{}.
\newblock \showarticletitle{Embedding Optimization for Training Large-scale Deep Learning Recommendation Systems with EMBark}. In \bibinfo{booktitle}{\emph{Proceedings of the 18th ACM Conference on Recommender Systems}}. \bibinfo{pages}{622--632}.
\newblock


\bibitem[Lyu et~al\mbox{.}(2020)]%
        {lyu2020deep}
\bibfield{author}{\bibinfo{person}{Ze Lyu}, \bibinfo{person}{Yu Dong}, \bibinfo{person}{Chengfu Huo}, {and} \bibinfo{person}{Weijun Ren}.} \bibinfo{year}{2020}\natexlab{}.
\newblock \showarticletitle{Deep match to rank model for personalized click-through rate prediction}. In \bibinfo{booktitle}{\emph{Proceedings of the AAAI Conference on Artificial Intelligence}}, Vol.~\bibinfo{volume}{34}. \bibinfo{pages}{156--163}.
\newblock


\bibitem[Mudigere et~al\mbox{.}(2022)]%
        {mudigere2022software}
\bibfield{author}{\bibinfo{person}{Dheevatsa Mudigere}, \bibinfo{person}{Yuchen Hao}, \bibinfo{person}{Jianyu Huang}, \bibinfo{person}{Zhihao Jia}, \bibinfo{person}{Andrew Tulloch}, \bibinfo{person}{Srinivas Sridharan}, \bibinfo{person}{Xing Liu}, \bibinfo{person}{Mustafa Ozdal}, \bibinfo{person}{Jade Nie}, \bibinfo{person}{Jongsoo Park}, {et~al\mbox{.}}} \bibinfo{year}{2022}\natexlab{}.
\newblock \showarticletitle{Software-hardware co-design for fast and scalable training of deep learning recommendation models}. In \bibinfo{booktitle}{\emph{Proceedings of the 49th Annual International Symposium on Computer Architecture}}. \bibinfo{pages}{993--1011}.
\newblock


\bibitem[Naumov et~al\mbox{.}(2020)]%
        {naumov2020deep}
\bibfield{author}{\bibinfo{person}{Maxim Naumov}, \bibinfo{person}{John Kim}, \bibinfo{person}{Dheevatsa Mudigere}, \bibinfo{person}{Srinivas Sridharan}, \bibinfo{person}{Xiaodong Wang}, \bibinfo{person}{Whitney Zhao}, \bibinfo{person}{Serhat Yilmaz}, \bibinfo{person}{Changkyu Kim}, \bibinfo{person}{Hector Yuen}, \bibinfo{person}{Mustafa Ozdal}, {et~al\mbox{.}}} \bibinfo{year}{2020}\natexlab{}.
\newblock \showarticletitle{Deep learning training in facebook data centers: Design of scale-up and scale-out systems}.
\newblock \bibinfo{journal}{\emph{arXiv preprint arXiv:2003.09518}} (\bibinfo{year}{2020}).
\newblock


\bibitem[Naumov et~al\mbox{.}(2019)]%
        {naumov2019deep}
\bibfield{author}{\bibinfo{person}{Maxim Naumov}, \bibinfo{person}{Dheevatsa Mudigere}, \bibinfo{person}{Hao-Jun~Michael Shi}, \bibinfo{person}{Jianyu Huang}, \bibinfo{person}{Narayanan Sundaraman}, \bibinfo{person}{Jongsoo Park}, \bibinfo{person}{Xiaodong Wang}, \bibinfo{person}{Udit Gupta}, \bibinfo{person}{Carole-Jean Wu}, \bibinfo{person}{Alisson~G Azzolini}, {et~al\mbox{.}}} \bibinfo{year}{2019}\natexlab{}.
\newblock \showarticletitle{Deep learning recommendation model for personalization and recommendation systems}.
\newblock \bibinfo{journal}{\emph{arXiv preprint arXiv:1906.00091}} (\bibinfo{year}{2019}).
\newblock


\bibitem[Singh et~al\mbox{.}(2024)]%
        {singh2024better}
\bibfield{author}{\bibinfo{person}{Anima Singh}, \bibinfo{person}{Trung Vu}, \bibinfo{person}{Nikhil Mehta}, \bibinfo{person}{Raghunandan Keshavan}, \bibinfo{person}{Maheswaran Sathiamoorthy}, \bibinfo{person}{Yilin Zheng}, \bibinfo{person}{Lichan Hong}, \bibinfo{person}{Lukasz Heldt}, \bibinfo{person}{Li Wei}, \bibinfo{person}{Devansh Tandon}, {et~al\mbox{.}}} \bibinfo{year}{2024}\natexlab{}.
\newblock \showarticletitle{Better generalization with semantic ids: A case study in ranking for recommendations}. In \bibinfo{booktitle}{\emph{Proceedings of the 18th ACM Conference on Recommender Systems}}. \bibinfo{pages}{1039--1044}.
\newblock


\bibitem[Stich(2019)]%
        {stich2019local}
\bibfield{author}{\bibinfo{person}{Sebastian~Urban Stich}.} \bibinfo{year}{2019}\natexlab{}.
\newblock \showarticletitle{Local SGD Converges Fast and Communicates Little}. In \bibinfo{booktitle}{\emph{ICLR 2019-International Conference on Learning Representations}}.
\newblock


\bibitem[Tieleman(2012)]%
        {tieleman2012lecture}
\bibfield{author}{\bibinfo{person}{Tijmen Tieleman}.} \bibinfo{year}{2012}\natexlab{}.
\newblock \showarticletitle{Lecture 6.5-rmsprop: Divide the gradient by a running average of its recent magnitude}.
\newblock \bibinfo{journal}{\emph{COURSERA: Neural networks for machine learning}} \bibinfo{volume}{4}, \bibinfo{number}{2} (\bibinfo{year}{2012}), \bibinfo{pages}{26}.
\newblock


\bibitem[Tsang and Ahle(2023)]%
        {tsang2023clustering}
\bibfield{author}{\bibinfo{person}{Henry Tsang} {and} \bibinfo{person}{Thomas Ahle}.} \bibinfo{year}{2023}\natexlab{}.
\newblock \showarticletitle{Clustering the sketch: dynamic compression for embedding tables}.
\newblock \bibinfo{journal}{\emph{Advances in Neural Information Processing Systems}}  \bibinfo{volume}{36} (\bibinfo{year}{2023}), \bibinfo{pages}{72155--72180}.
\newblock


\bibitem[Xu et~al\mbox{.}(2020)]%
        {xu2020automatic}
\bibfield{author}{\bibinfo{person}{Yuanzhong Xu}, \bibinfo{person}{HyoukJoong Lee}, \bibinfo{person}{Dehao Chen}, \bibinfo{person}{Hongjun Choi}, \bibinfo{person}{Blake Hechtman}, {and} \bibinfo{person}{Shibo Wang}.} \bibinfo{year}{2020}\natexlab{}.
\newblock \showarticletitle{Automatic cross-replica sharding of weight update in data-parallel training}.
\newblock \bibinfo{journal}{\emph{arXiv preprint arXiv:2004.13336}} (\bibinfo{year}{2020}).
\newblock


\bibitem[Yin et~al\mbox{.}(2021)]%
        {yin2021tt}
\bibfield{author}{\bibinfo{person}{Chunxing Yin}, \bibinfo{person}{Bilge Acun}, \bibinfo{person}{Carole-Jean Wu}, {and} \bibinfo{person}{Xing Liu}.} \bibinfo{year}{2021}\natexlab{}.
\newblock \showarticletitle{Tt-rec: Tensor train compression for deep learning recommendation models}.
\newblock \bibinfo{journal}{\emph{Proceedings of Machine Learning and Systems}}  \bibinfo{volume}{3} (\bibinfo{year}{2021}), \bibinfo{pages}{448--462}.
\newblock


\bibitem[Zeng et~al\mbox{.}(2024)]%
        {zeng2024interformer}
\bibfield{author}{\bibinfo{person}{Zhichen Zeng}, \bibinfo{person}{Xiaolong Liu}, \bibinfo{person}{Mengyue Hang}, \bibinfo{person}{Xiaoyi Liu}, \bibinfo{person}{Qinghai Zhou}, \bibinfo{person}{Chaofei Yang}, \bibinfo{person}{Yiqun Liu}, \bibinfo{person}{Yichen Ruan}, \bibinfo{person}{Laming Chen}, \bibinfo{person}{Yuxin Chen}, {et~al\mbox{.}}} \bibinfo{year}{2024}\natexlab{}.
\newblock \showarticletitle{InterFormer: Towards Effective Heterogeneous Interaction Learning for Click-Through Rate Prediction}.
\newblock \bibinfo{journal}{\emph{arXiv preprint arXiv:2411.09852}} (\bibinfo{year}{2024}).
\newblock


\bibitem[Zha et~al\mbox{.}(2022a)]%
        {zha2022autoshard}
\bibfield{author}{\bibinfo{person}{Daochen Zha}, \bibinfo{person}{Louis Feng}, \bibinfo{person}{Bhargav Bhushanam}, \bibinfo{person}{Dhruv Choudhary}, \bibinfo{person}{Jade Nie}, \bibinfo{person}{Yuandong Tian}, \bibinfo{person}{Jay Chae}, \bibinfo{person}{Yinbin Ma}, \bibinfo{person}{Arun Kejariwal}, {and} \bibinfo{person}{Xia Hu}.} \bibinfo{year}{2022}\natexlab{a}.
\newblock \showarticletitle{Autoshard: Automated embedding table sharding for recommender systems}. In \bibinfo{booktitle}{\emph{Proceedings of the 28th ACM SIGKDD Conference on Knowledge Discovery and Data Mining}}. \bibinfo{pages}{4461--4471}.
\newblock


\bibitem[Zha et~al\mbox{.}(2023)]%
        {zha2023pre}
\bibfield{author}{\bibinfo{person}{Daochen Zha}, \bibinfo{person}{Louis Feng}, \bibinfo{person}{Liang Luo}, \bibinfo{person}{Bhargav Bhushanam}, \bibinfo{person}{Zirui Liu}, \bibinfo{person}{Yusuo Hu}, \bibinfo{person}{Jade Nie}, \bibinfo{person}{Yuzhen Huang}, \bibinfo{person}{Yuandong Tian}, \bibinfo{person}{Arun Kejariwal}, {et~al\mbox{.}}} \bibinfo{year}{2023}\natexlab{}.
\newblock \showarticletitle{Pre-train and search: Efficient embedding table sharding with pre-trained neural cost models}.
\newblock \bibinfo{journal}{\emph{Proceedings of Machine Learning and Systems}}  \bibinfo{volume}{5} (\bibinfo{year}{2023}), \bibinfo{pages}{68--88}.
\newblock


\bibitem[Zha et~al\mbox{.}(2022b)]%
        {zha2022dreamshard}
\bibfield{author}{\bibinfo{person}{Daochen Zha}, \bibinfo{person}{Louis Feng}, \bibinfo{person}{Qiaoyu Tan}, \bibinfo{person}{Zirui Liu}, \bibinfo{person}{Kwei-Herng Lai}, \bibinfo{person}{Bhargav Bhushanam}, \bibinfo{person}{Yuandong Tian}, \bibinfo{person}{Arun Kejariwal}, {and} \bibinfo{person}{Xia Hu}.} \bibinfo{year}{2022}\natexlab{b}.
\newblock \showarticletitle{Dreamshard: Generalizable embedding table placement for recommender systems}.
\newblock \bibinfo{journal}{\emph{Advances in Neural Information Processing Systems}}  \bibinfo{volume}{35} (\bibinfo{year}{2022}), \bibinfo{pages}{15190--15203}.
\newblock


\bibitem[Zhai et~al\mbox{.}(2024)]%
        {zhaiactions}
\bibfield{author}{\bibinfo{person}{Jiaqi Zhai}, \bibinfo{person}{Lucy Liao}, \bibinfo{person}{Xing Liu}, \bibinfo{person}{Yueming Wang}, \bibinfo{person}{Rui Li}, \bibinfo{person}{Xuan Cao}, \bibinfo{person}{Leon Gao}, \bibinfo{person}{Zhaojie Gong}, \bibinfo{person}{Fangda Gu}, \bibinfo{person}{Jiayuan He}, {et~al\mbox{.}}} \bibinfo{year}{2024}\natexlab{}.
\newblock \showarticletitle{Actions Speak Louder than Words: Trillion-Parameter Sequential Transducers for Generative Recommendations}. In \bibinfo{booktitle}{\emph{Forty-first International Conference on Machine Learning}}.
\newblock


\bibitem[Zhang et~al\mbox{.}(2024b)]%
        {zhang2024wukong}
\bibfield{author}{\bibinfo{person}{Buyun Zhang}, \bibinfo{person}{Liang Luo}, \bibinfo{person}{Yuxin Chen}, \bibinfo{person}{Jade Nie}, \bibinfo{person}{Xi Liu}, \bibinfo{person}{Shen Li}, \bibinfo{person}{Yanli Zhao}, \bibinfo{person}{Yuchen Hao}, \bibinfo{person}{Yantao Yao}, \bibinfo{person}{Ellie~Dingqiao Wen}, {et~al\mbox{.}}} \bibinfo{year}{2024}\natexlab{b}.
\newblock \showarticletitle{Wukong: towards a scaling law for large-scale recommendation}. In \bibinfo{booktitle}{\emph{Proceedings of the 41st International Conference on Machine Learning}}. \bibinfo{pages}{59421--59434}.
\newblock


\bibitem[Zhang et~al\mbox{.}(2022)]%
        {zhang2022dhen}
\bibfield{author}{\bibinfo{person}{Buyun Zhang}, \bibinfo{person}{Liang Luo}, \bibinfo{person}{Xi Liu}, \bibinfo{person}{Jay Li}, \bibinfo{person}{Zeliang Chen}, \bibinfo{person}{Weilin Zhang}, \bibinfo{person}{Xiaohan Wei}, \bibinfo{person}{Yuchen Hao}, \bibinfo{person}{Michael Tsang}, \bibinfo{person}{Wenjun Wang}, {et~al\mbox{.}}} \bibinfo{year}{2022}\natexlab{}.
\newblock \showarticletitle{DHEN: A deep and hierarchical ensemble network for large-scale click-through rate prediction}.
\newblock \bibinfo{journal}{\emph{arXiv preprint arXiv:2203.11014}} (\bibinfo{year}{2022}).
\newblock


\bibitem[Zhang et~al\mbox{.}(2024a)]%
        {zhang2024scaling}
\bibfield{author}{\bibinfo{person}{Gaowei Zhang}, \bibinfo{person}{Yupeng Hou}, \bibinfo{person}{Hongyu Lu}, \bibinfo{person}{Yu Chen}, \bibinfo{person}{Wayne~Xin Zhao}, {and} \bibinfo{person}{Ji-Rong Wen}.} \bibinfo{year}{2024}\natexlab{a}.
\newblock \showarticletitle{Scaling law of large sequential recommendation models}. In \bibinfo{booktitle}{\emph{Proceedings of the 18th ACM Conference on Recommender Systems}}. \bibinfo{pages}{444--453}.
\newblock


\bibitem[Zhang et~al\mbox{.}(2019)]%
        {zhang2019deep}
\bibfield{author}{\bibinfo{person}{Shuai Zhang}, \bibinfo{person}{Lina Yao}, \bibinfo{person}{Aixin Sun}, {and} \bibinfo{person}{Yi Tay}.} \bibinfo{year}{2019}\natexlab{}.
\newblock \showarticletitle{Deep learning based recommender system: A survey and new perspectives}.
\newblock \bibinfo{journal}{\emph{ACM computing surveys (CSUR)}} \bibinfo{volume}{52}, \bibinfo{number}{1} (\bibinfo{year}{2019}), \bibinfo{pages}{1--38}.
\newblock


\bibitem[Zhao et~al\mbox{.}(2020)]%
        {zhao2020distributed}
\bibfield{author}{\bibinfo{person}{Weijie Zhao}, \bibinfo{person}{Deping Xie}, \bibinfo{person}{Ronglai Jia}, \bibinfo{person}{Yulei Qian}, \bibinfo{person}{Ruiquan Ding}, \bibinfo{person}{Mingming Sun}, {and} \bibinfo{person}{Ping Li}.} \bibinfo{year}{2020}\natexlab{}.
\newblock \showarticletitle{Distributed hierarchical gpu parameter server for massive scale deep learning ads systems}.
\newblock \bibinfo{journal}{\emph{Proceedings of Machine Learning and Systems}}  \bibinfo{volume}{2} (\bibinfo{year}{2020}), \bibinfo{pages}{412--428}.
\newblock


\bibitem[Zhao et~al\mbox{.}(2023)]%
        {zhao2023pytorch}
\bibfield{author}{\bibinfo{person}{Yanli Zhao}, \bibinfo{person}{Andrew Gu}, \bibinfo{person}{Rohan Varma}, \bibinfo{person}{Liang Luo}, \bibinfo{person}{Chien-Chin Huang}, \bibinfo{person}{Min Xu}, \bibinfo{person}{Less Wright}, \bibinfo{person}{Hamid Shojanazeri}, \bibinfo{person}{Myle Ott}, \bibinfo{person}{Sam Shleifer}, {et~al\mbox{.}}} \bibinfo{year}{2023}\natexlab{}.
\newblock \showarticletitle{Pytorch fsdp: experiences on scaling fully sharded data parallel}.
\newblock \bibinfo{journal}{\emph{arXiv preprint arXiv:2304.11277}} (\bibinfo{year}{2023}).
\newblock


\end{thebibliography}

% \clearpage

% \appendix

% \input{Appendix/appedix}

\end{document}